# On the Interaction in Lattice Gauge Theory

Bijan Sheikholeslami-Sabzevari[1], Hamideh Rahmati[2]

[1]-Department of physics, University of Isfahan, Hezar Gerib Ave., 81746 Isfahan, Iran.

b_sh.sabzevari@yahoo.com

2-Department of Physics, University of Guilan, Rasht, Iran, P.O.Box:41335-1915.

hrahmati1357@yahoo.com

## Abstract

Haag`s theorem states that if a quantum field theory is Lorentz invariant and irreducible, there is no interaction picture. But if we construct quantum field theory on a discrete lattice space-time, it's representation will be reducible and the interaction picture will be restored again.

# 1 Introduction

Quantum field theory deals with systems that have infinite degrees of freedom. This theory is a frame work for relativistic quantum mechanical models of field-like or many-body systems and has much success in solving problems concerning elementary particles. But it encounters some difficulties caused by the nonlinear field structure, occurring in coupled equations and Green's functions, products of distributions, etc…which have no precise mathematical and physical solutions [1].

There are various approaches to the study of quantum fields. One method is stated by Haag and Kastler that introduces axiomatic quantum field theory [2]. The problem in this approach is called Haag`s theorem which states that there is no interaction picture. It is proved by this theorem that non-equivalent canonical commutation relations are posed and different vacuums appear that aren't meaningful. With the result that one can show that every interacting quantum field theory (QFT) is equivalent to a free field theory, i.e., there are no interactions in elementary particle physics. This is shown by the non-existent of the interaction picture. On the other side, since the only way to calculate cross sections is by using perturbation theory via the interaction picture and these cross sections actually can be and are measured and observed, we come in conflict with the foundations of QFT. In short: Haag`s theorem states that there is not any interaction as a start for perturbation theory in QFT. Van Hove, Segal, Araki, Wightman and Guenin have circumvented this theorem in different ways [3, 4].

Here we introduce the approach of lattice gauge theory. This theory examines quantum fields in discrete space-time and was introduced by K.Wilson in 1974 being very successful in solving problems of quantum fields in recent years [5, 6]. In this article Haag`s theorem is circumvented by using lattice gauge theory.

# 2 Haag's theorem and its rejections

The axiomatic quantum field theory used by Haag is as follows:
If $\vartheta$ is an open region in Minkowski space and the algebra of observables in this region is $A(\vartheta)$, we have the following correspondence:

$$\vartheta \to A(\vartheta)$$

This mathematical frame should have the following properties (for a more detailed outline see the appendix):

*1-$\vartheta$* Regions for which the correspondence $\vartheta \to A(\vartheta)$ is defined are the open sets with compact closure in Minkowski space (compact closure means finite extension of sets).

2-Isotony: If $\vartheta_1$ includes $\vartheta_2$, then the $A(\vartheta_1)$ algebra also includes the $A(\vartheta_2)$ algebra, i.e., $\vartheta_1 \supset \vartheta_2 \Rightarrow A(\vartheta_1) \supset A(\vartheta_2)$. $A(\vartheta_1)$ and $A(\vartheta_2)$ have a common unit element or neither of them has a unit element.

3-Local commutatively: If $\vartheta_1$ and $\vartheta_2$ are completely space-like, then $A(\vartheta_1)$ and $A(\vartheta_2)$ commute. In other words open space-like regions have independent observables in Minkowski space.

4-The union of all $A(\vartheta)$ is a normed *-algebra.

5-Lorentz invariance: The inhomogeneous Lorentz group is represented by the following authomorphisms:
$$a \in A \Rightarrow a^L \in A, A(\vartheta) \xrightarrow{L} A(L\vartheta)$$
Where $L\vartheta$ is the image of $\vartheta$ under Lorentz transformations $L$.

6-*A* is primitive; i.e., it accepts a faithful irreducible representation, i.e., or for any operator only one representation [2].

In axioms (5) and (6) only one-to-one images are relevant. But a one-to-one property is not established in interacting fields. Interaction terms have powers greater than one and are not linear; so by working with single states, they result in other states and then nonequivalent or strange representations appear. If this is adopted, the theory gains much in richness, but simultaneously, the mathematical structure becomes more involved. Each particular representation will be associated with a particular Hilbert space. The uniqueness of the vacuum is, of course, valid only in one particular representation. In other words, when the states of the system are not invariant, nonequivalent representations and therefore canonical and non-simultaneous commutation relations appear [1, p.388-394].

One of the basic assumptions used by Haag to prove his theorem is that "equal time commutation relations are established for physical fields and there are no nonequivalent representations of the system" [1, p.388]. An appropriate solution to this problem is that if we consider the theory on a space time lattice, explicit answers are found because the theory is automaticaly renormalized, therefore a natural cut off appears in measuring energy, position, etc.

Van Hove found that Hamiltonians will be well-defined if they are expressed in the form of Hermitian operators which operate only under very tiny subsets which are orthogonal vectors of non-separable Hilbert space. Segal showed that special classes of interacting Hamiltonians can be mathematically meaningful in appropriate strange representations .Araki considered that the Hamiltonian of field theory is unique under special conditions and can satisfy canonical commutation relations. Wightman stated that there is a difference between adjoint and hermitian operators. An adjoint operator concludes unique dynamics but hermitian operator may conclude none or some dynamics (strange and nonequivalent representations). His purpose was that interaction operators should be defined in the adjoint form to give a unique and hence a real interaction representation [3].

Guenin rejected Haag`s theorem by presenting a new picture (very similar to Dirac`s). As mentioned, Haag`s theorem states that if Lorentz invariance is established in the theory, Hamiltonian interaction isn't unique, but Guenin showed that this subject couldn't prove the lack of a map from *-algebra to operator rings at different times. He showed that there is an isomorphic map and it is unique, localized and meaningful in the Hilbert space. The advantage of the Guenin interaction picture is that it operates in Fock space. In the Guenin method it is possible to solve Haag`s theorem by violating Lorentz invariance and by limiting the system in a finite box [4].

## 3 Rejecting Haag's theorem on discrete space-time lattice

QFT on the lattice in the framework of non-perturbative theory is very successful in describing QED, QCD and weak interactions to a relative high precession, from first principles. An apparent mathematical inconsistency in QED is the existence of the so-called triviality problem (first discovered by L.D.Landau and collaborators [10]) in the perturbative behavior of the renormalized coupling constant at high energy [11]. The renormalization group provides useful insight in this case. The Callan-Symanzik β-function is defined as [6]:

$$\beta(g_\mu) = \frac{\mu dg_\mu}{d\mu} \qquad (1)$$

Here 'g' is the coupling constant and 'μ' is an arbitrary mass providing a renormalization scale. In the 1-loop approximation for $\mu_{Landau}$ we have:

$$\mu_{Landau} = \mu_0 \exp\left(\frac{4\pi}{\beta_1 e_R^2(\mu_0)}\right) \qquad (2)$$

Substituting here the physical value of the renormalized coupling constant ($e_R^2(\mu_0) = \frac{4\pi}{137}$ and $\beta_1 = \frac{2}{3\pi}$), one obtains a very very high scale ($\mu_{Landau} \propto \exp(645.27)$). Therefore, in QED the physical inconsistent energy range is very far away from any reasonable scale.

To avoid Haag's theorem, a four corners of the lattice fields are set up and there are gauge fields between them causing transformation from one point of the lattice to the other point. These fields are members of the SU (N) group [7]. According to the theorem (3.2.2) in [8]: "any representation of finite group, is equivalent to a representation by unitary services"; so any finite representation of groups on Wilson lattice are unitary and reducible.

As mentioned above Haag's theorem has shown that the transformation between interacting and free field operators cannot be unitary. Now if the Haag-Kastler net is spread over a finite lattice of space-time, it has finite representations of unitary groups. For finite lattice spacing the theory involves interaction, because g≠0. In particular, long before we reach the energies where QED becomes trivial, it seems necessary to take also gravitation into account, and yet no one knows how to calculate the effects of strong gravitational forces at such energies! That Haag's theorem will not work on the lattice is expected, because Lorentz invariance is broken on a discrete space-time.

## 4 Conclusions

The success of relativistic quantum field theory in calculating physical quantities in the experimental and theoretical domain led physicist to a logical puzzle: is QFT mathematically consistent? Can one give a mathematically complete example of any non-linear theory relevant for the description of interacting particles whose solutions incorporate relativistic covariance, positive energy and causality?

There is today a view that interacting QFTs are factually trivial and hence not mathematically consistent. There are theorems [9, 11] showing that in general this theory does not have an interacting continuum limit unless the theory is asymptotically free. But before we reach such extreme high energies, we should know how to calculate the effects of strong gravitational fields.

In this paper we considered Haag's theorem and it was rejected by constructing QFT on discrete lattice space time. There must be a relation between Haag's theorem, non-interaction, Lorentz invariance, lattice gauge theory and continuum limit, which is not clarified yet. Further investigations on this problem are on the way.

**Appendix**

In the early 1960s, an alternative version of the foundations of relativistic quantum field theory was developed by Haag and Kastler, based on the use of algebras of bounded operators [2]. The basic objects of the local algebraic formalism is a C*-algebra and operators associated with bounded open sets in space-time.

The set $A$ is named an *-algebra if for any member $a \in A$ there is a conjugate member a* as follows:

$$1:(a^*)^* = a \;\; ; 2:(ab)^* = b^*a^* \;\; ; 3:(\lambda a)^* = \bar{\lambda}a^*, \lambda \in C \;\; ; 4:(a+b)^* = a^*+b^*$$

If for any member "a" there exist a real number $\|a\|$ by the following conditions, the set A is named a normed algebra:

$$1 - \|a\| \geq 0, \|a\| = 0 \Leftrightarrow a = 0 \;\; ; 2 - \|\lambda a\| = |\lambda|\|a\| \;\; ; 3 - \|a+b\| \leq \|a\| + \|b\|, \|ab\| \leq \|a\|\|b\|$$

If the algebra on a vector space H has members with complete norm, it is named a Banach algebra and if Banach *-algebra has the condition $\|a^*a\| = \|a\|^2$ too, the algebra is a C*-algebra.

For the computational purposes of quantum theory it is sufficient to consider only those representations of a C*-algebra (which in quantum theory is the algebra generated by the observables) that are irreducible and form a complete set of operators i.e. a ring. Segal pointed out that many questions of physical interest (e.g. the determination of spectral values) can be answered without reference to the Hilbert space if one chooses the algebra of observables to be a C*-algebra[1]. Von Neumann's theorem is an important theorem in quantum field theory concerning Haag's theorem. This theorem is as follows:

Suppose (a) a field $\varphi(x)$ obeys the ETCR[2] :

$$[\varphi(x), \varphi(x')] = 0 \;\; ;[\dot{\varphi}(x), \dot{\varphi}(x')] = 0 \;\; ;[\varphi(x), \dot{\varphi}(x')] = \delta(\vec{x} - \vec{x}') \qquad \text{and}$$

(b) this algebra has no inequivalent representations, then the field system $\varphi(\vec{x})$ (at any fixed time $x_0 = t$ ) is an irreducible operator ring. The converse theorem also holds true: if

---

1- Segal, I. E., Bulletin of the American Mathematical Society, 73, (1947) 53.
2- Equal Time Commutation Relations.

(a) the ETCR applies and if (b) the operator ring is irreducible, then all representations of the algebra are unitary equivalent [1, page 330].

The C*-algebra defined by Haag and Kastler could not determine how the phenomena associated with unitary inequivalent representations of the algebras of observables would appear in concrete Lagrangian field theories[1].

---

1- Streater, R.F., Wightman, A.S.: PCT, Spin Statistics and All That, Addison-Wesley (1964).